\documentclass[12pt,onecolumn,floatfix,aps,prd,showpacs,amsmath,nofootinbib]{revtex4}
\usepackage{times}
\usepackage{color}
\usepackage{graphicx}
\usepackage{xcomment}
\usepackage{ulem}
\newcommand{\sfig}[2]{
\includegraphics[width=#2]{#1}
        }

\newcommand{\Sfig}[2]{
    \begin{figure}[thbp]
    \sfig{#1.eps}{.5\columnwidth}
    \caption{{\small #2}}
    \label{fig:#1}
    \end{figure}
}

\newcommand{\rf}[1]{\ref{fig:#1}}

\def\knew{\kappa_{\rm ln}}

\def\lsim{\mathrel{\raise.3ex\hbox{$<$\kern-.75em\lower1ex\hbox{$\sim$}}}}
\def\gsim{\mathrel{\raise.3ex\hbox{$>$\kern-.75em\lower1ex\hbox{$\sim$}}}}

\def\cmm2{{\,\rm cm^{-2}}}
\def\cm2{{\,{\rm cm}^2}}
\def\cmm3{{\,{\rm cm}^{-3}}}
\def\gcmm3{{\,{\rm g\,cm^{-3}}}}

\def\fun#1#2{\lower3.6pt\vbox{\baselineskip0pt\lineskip.9pt
  \ialign{$\mathsurround=0pt#1\hfil##\hfil$\crcr#2\crcr\sim\crcr}}}

\def\be{\begin{equation}}
\def\ee{\end{equation}}
\def\bea{\begin{eqnarray}}
\def\eea{\end{eqnarray}}

\def\be{\begin{equation}}
\def\ee{\end{equation}}
\def\bea{\begin{eqnarray}}
\def\eea{\end{eqnarray}}

\begin{document}

\title{Re-capturing cosmic information}

\author{Hee-Jong Seo$^1$, Masanori Sato$^2$, Scott Dodelson$^{1,3,4}$, Bhuvnesh Jain$^5$, and Masahiro Takada$^6$}
%\email{$^1$avalli@fnal.gov}
%\email{$^2$dodelson@fnal.gov}
%\email{$^4$zhangpengjie@gmail.com}
\affiliation{$^1$Center for Particle Astrophysics, Fermi National
Accelerator Laboratory, Batavia, IL~~60510}
\affiliation{$^2$Department of Physics, Nagoya University, Nagoya 464-8602, Japan}
\affiliation{$^3$Department of Astronomy \& Astrophysics, The
University of Chicago, Chicago, IL~~60637}
\affiliation{$^4$Kavli Institute for Cosmological Physics, Chicago, IL~~60637}
\affiliation{$^5$Department of Physics \& Astronomy, Center for
  Particle-Cosmology, University of Pennsylvania, Philadelphia PA 19104}
\affiliation{$^6$Institute for the Physics and Mathematics of the
Universe (IPMU), University of Tokyo, Chiba 277-8582, Japan}
\date{\today}

\begin{abstract}
Gravitational lensing of distant galaxies 
can be exploited to infer the {\it convergence} field as a function of angular position on the sky. 
The statistics of this field, much like that of the cosmic microwave
 background (CMB),  can be studied to extract information about fundamental
 parameters in cosmology, most notably the dark energy in the
 Universe. Unlike the CMB, the distribution of matter in the Universe
 which determines the convergence field is highly non-Gaussian,
 reflecting the nonlinear processes which accompanied structure
 formation. Much of the cosmic information contained in the
 initial  field is therefore unavailable to the standard power
 spectrum measurements.  Here we propose a method for re-capturing 
cosmic information by using the power spectrum of a simple function of the observed (nonlinear) convergence field. 
We adapt the approach of Neyrinck et al. (2009) to lensing by using a 
modified logarithmic transform of the convergence field. The
Fourier transform of the log-transformed field has modes that are 
nearly uncorrelated, which allows for additional cosmological
information to be extracted from small-scale modes. 
\end{abstract}

\newcommand\est{\Delta}
\newcommand\kz{\kappa_0}

\maketitle

\section{Introduction}
Gravitational lensing has emerged as a powerful tool to probe the distribution of matter in the Universe~\cite{Bartelmann:1999yn}. Observations of the ellipticities of
background galaxies can be transformed into estimates of the {\it
convergence} field $\kappa(\vec\theta)$. Along a given line of sight
$\vec\theta$, the convergence measures a weighted integral of the total mass 
density field. Thus by carefully studying $\kappa$ as a function of position on the sky, we can learn about the underlying density field directly, without relying on the traditional assumption that every galaxy corresponds to an overdense region.

By measuring the convergence to sources at multiple background
redshifts, cosmologists can infer not only the density field as a
function of 2D position~\cite{Kaiser:1991qi,Blandford:1991zz,Jain:1996st,Jain:1999ir}, but also the evolution of this density field
with time~\cite{Hu:1999ek}. This information will be particularly valuable as a tool to
study both dark matter and dark energy, which affect the growth of
structure in the Universe~\cite{Huterer:2001yu,Hu:2001fb}. A number of wide-area
surveys have been planned with the goal of mapping out the cosmic convergence field, and ultimately measuring properties of the dark energy~\cite{Aldering:2005qn,2006SPIE.6269E...9M,Tyson:2006hs,Abbott:2005bi,Refregier:2008js}.

This goal appears attainable as it is reminiscent of another cosmological success story: measurement of anisotropies in the CMB~\cite{Hu:2001bc}. In both cases, the values of the measured quantities -- temperature in the case of the CMB and convergence from lensing -- at any particular spot on the sky are not important. Rather, it is the statistics of the field that carries all the important information. The two-point function of the temperature of the CMB, the {\it power spectrum} of the anisotropies, is sensitive to a number of cosmological parameters, and some of these have now been measured to percent level accuracy~\cite{Komatsu:2010fb}. Similarly, the power spectrum of the convergence depends on cosmological parameters, and one can hope to extract information about these parameters from lensing surveys~\cite{Hu:1998az,Abazajian:2002ck,Refregier:2003xe,Hoekstra:2008db}. 

However the convergence field differs in an important way
from the anisotropy maps. CMB anisotropies provide a snapshot of the 
Universe when it was very young, and hence all deviations from homogeneity 
are very small (temperature differences in the maps are of order several
parts in a hundred thousand). The physics describing these perturbations
is linear. Further, the perturbations were drawn from a Gaussian
distribution, so the two-point function captures all of the information
in the field. On the other hand, the cosmic density field today is
non-linear and non-Gaussian, increasingly so on smaller scales,
so some of the information initially stored in the two-point function when the fields were linear is no longer present. 

Before quantifying this notion that information has left the two-point function, it is worthwhile to review some approaches to this problem. 
Takada \& Jain~\cite{Takada:2003ef} pointed out that including information
from both the two- and three-point functions significantly reduces the
errors on cosmological parameters. This makes intuitive sense: the
nonlinear process of gravitational instability transforms the initially
Gaussian field into one with appreciable non-Gaussianity, one hallmark
of which is a non-zero skewness. The goal of
measuring both sets of functions may work, but it suffers from the
drawback of requiring non-trivial covariance matrices (which involve the challenge of computing five- and six-point functions) \cite{TakadaJain:09}. 

A series of papers devoted to the 3D density field $\delta(\vec x)\equiv (\rho(\vec x)-\bar\rho)/\bar\rho$ ~\cite{Rimes:2005xs,Rimes:2005dz,Neyrinck:2006zi,Neyrinck:2006xd,Neyrinck:2009fs}
have noted that information in the power spectrum of $\delta$ saturates at high wavenumbers $k$ 
(or small length scales). That is, the power spectrum at high-$k$ is highly correlated, apparently due to the coupling of modes induced by nonlinear gravitational clustering. 
 The most recent of these papers offered a useful proposal~\cite{Neyrinck:2009fs}
for re-capturing information about the 3D density field by pointing out
that $\ln(1+\delta)$ has 
 properties similar to
the initial, {\it linear} density field. Its probability distribution is close to a Gaussian, 
the broadband shape is close to that of the linear power spectrum, and
finally, the information content is close to the Gaussian case. Practically this transform may be of limited utility because
the 3D density field is typically estimated by using galaxies as tracers, and  it is unlikely that the log transform of the {\it galaxy}
density will be a useful tracer of the linear {\it matter} density field.
 However, we now show that the log transform can be applied to the 2D lensing convergence field to de-correlate modes and obtain  information from 
higher-order correlations back in the two-point function.

\section{Log-mapping for lensing}

Using simulations, we study the
statistics of a new field:
\begin{equation}
\knew(\vec\theta) \equiv {\kappa_0}\ln\left[ 1 + \frac{\kappa(\vec\theta)}{\kappa_0}\right]
\end{equation}
where $\kappa_0$ is a constant with a value slightly larger than the
absolute value of the minimum value of $\kappa$ in the survey -- this  
keeps the argument of the logarithm positive.
 In the limit of small
$\kappa$, $\knew$ reduces to the standard convergence, but the log alters 
it in very high or low density regimes.
The parameter $\kappa_0$ tunes the degree of the alteration: the smaller $\kappa_0$, the more we alter the field\footnote{For our fiducial maps with 0.15 arcmin pixel scale, we use $\kappa_0=0.0482$ based on the minimum value of measured $\kappa$. }.
The log-mapping described above is motivated by our goal to de-correlate the Fourier modes 
of the convergence field. Although the mapping is local on the sky, it is nonlinear, so in Fourier 
space it has the potential to undo some of the correlations introduced by nonlinear clustering. 

\Sfig{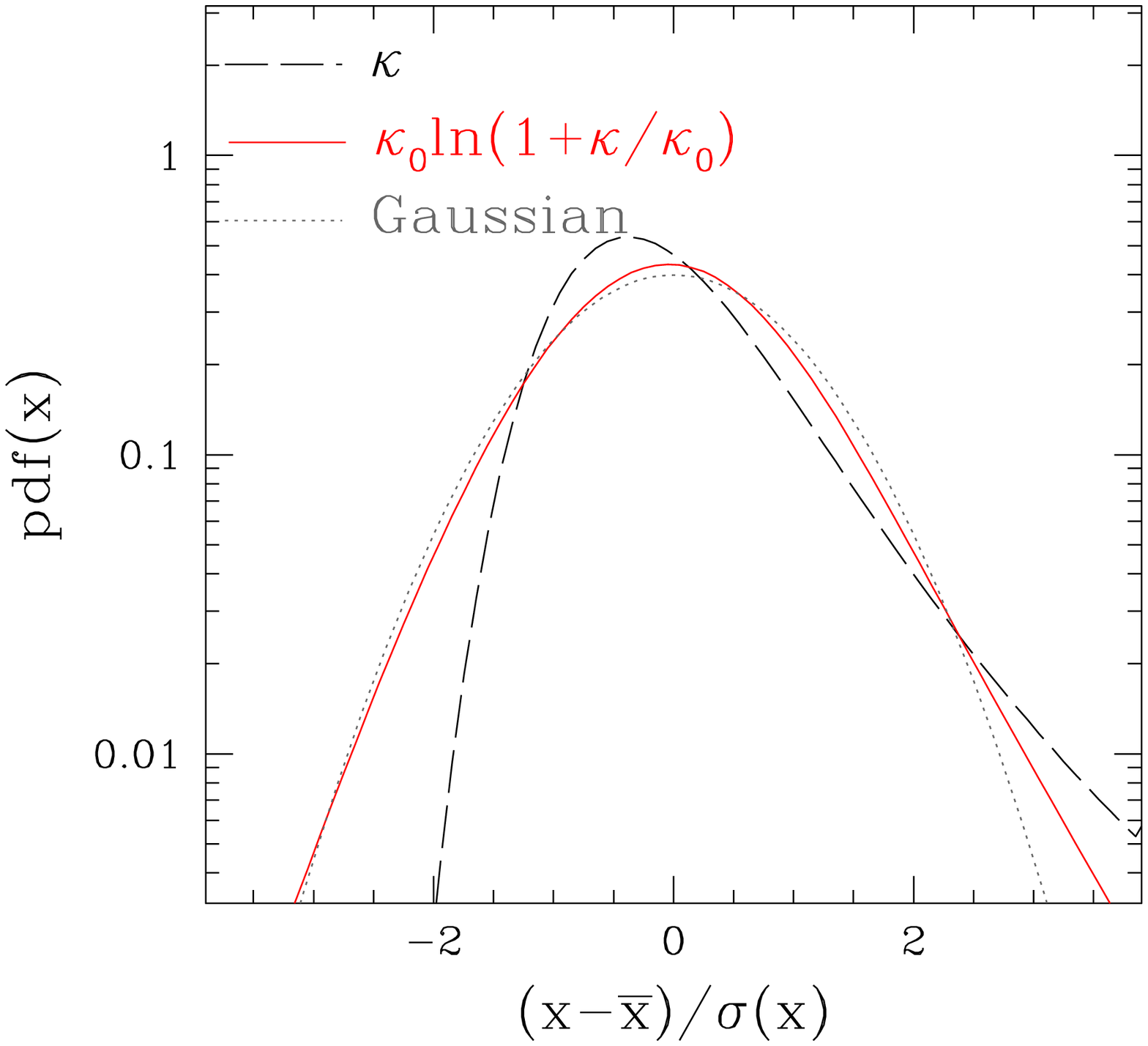}{The probability distribution function of the  two
fields $\kappa$ (black, dashed) and $\knew$ (red, solid) in comparison
with a Gaussian Probability Distribution Function (dotted). The skewed PDF of $\kappa$ reflects the distribution of structure in the Universe: large underdense regions separated by some very overdense regions. The log transform restores the field to a PDF that is nearly Gaussian.}

 To study the properties of $\knew$, we use a suite of numerical simulations: 100 
 convergence fields, each $5^\circ \times 5^\circ$ (a total of 2500
square degrees) with $2048^2$ pixels (i.e., 0.15 arcmin per pixel) 
were generated using N-body simulations as described in
\cite{Sato:2009ct}.  All source galaxies are taken to be at redshift 
$z_s=1$
for all the results shown below, 
though we have also checked other source redshifts.

A first glimpse into the advantages of the log transform can be seen from
Fig.~\rf{pdf} which shows the probability distribution function (PDF) of
both $\kappa$ and $\knew$, 
compared to the (linear) Gaussian PDF. The new field is much closer to Gaussian, a promising sign since the loss of information in $\kappa$ is attributed to gravity transforming the initially Gaussian 
random field into one that is highly non-Gaussian.

\Sfig{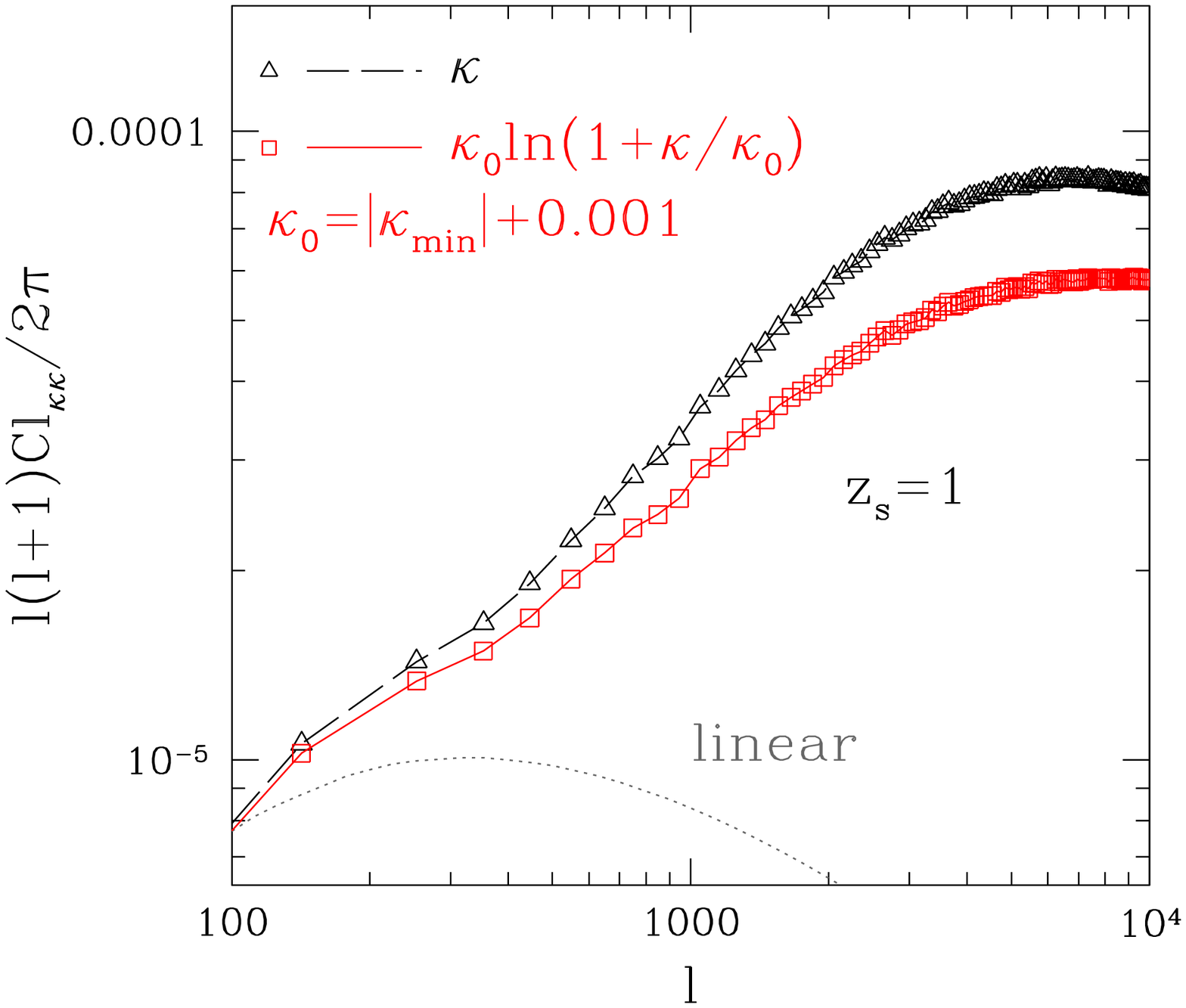}{The measured power spectra of the convergence field $\kappa$ and the log transformed 
field $\knew$. The latter has smaller amplitude at high $l$. The linear power spectrum is shown 
by the dotted curve. }

\def\lmax{l_{\rm max}}
\def\cov{{\rm Cov}}

To evaluate the log transform quantitatively, we take the Fourier transform of
the three different  convergence
 fields (linear, $\kappa$, and
$\knew$) in each of the simulations. The angular power spectrum is estimated from 
the Fourier transforms (denoted $\tilde\kappa(\vec
l)$) by summing over all modes
with wavenumber $\vert\vec l\vert$ in a given bin $l_{\rm bin}$.

Fig.~\rf{cl} shows these spectra. As expected, the power spectrum of the
nonlinear $\kappa$ field is much larger than the linear field on small
scales (large $l$). This excess power on small scales is suppressed when $\knew$ is used. Again the result is not surprising, as the high density regions are smoothed out: $\knew \ll \kappa$ for large $\kappa$.

\section{Recovery of cosmological information}

Although the power spectrum of $\knew$ is smaller than that of $\kappa$, it contains more cosmological information. 
To see this, consider a model with one free parameter, the 
amplitude of the {\it observed, nonlinear}
 power spectrum before and after the log transform.
 The projected fractional error on this parameter is the inverse of the signal to noise  defined as
\begin{equation}
\frac{S}{N}(\lmax) \equiv \left[ \sum_{l,l'<\lmax} C_l \cov^{-1}(l,l') C_{l'} \right]^{1/2}
\end{equation}
where $C_l$ is the power spectrum of multipole $l$  before and after the transform, $\cov$ is the covariance matrix describing correlations between 
the power spectra of multipoles $l$ and $l'$ ($l,l' < \lmax$),
 and the summation runs over all the
multipoles $l$ and $l'$ subject to $l,l'<\lmax$ \cite{Sato:2009ct,Takahashi2009}. 
We follow \cite{Neyrinck:2009fs} and call the square of the $S/N$ ratio the 
information content. Heuristically, then, ``information'' quantifies how accurately parameters will be determined. 
To compute 
the expected error on the chosen cosmological parameter
(here the amplitude of the power
 spectrum \cite{Lee2008}),
 one needs to know the
covariance matrix of the spectra. If the field was Gaussian 
random, the covariance matrix would be diagonal. In the absence of shape noise\footnote{The ellipticity of a single galaxy is, in the absence of any distortion by the intervening density field, randomly distributed on the sky with an RMS of about 0.3. This corresponds to noise in the measurement of the cosmic convergence field, a noise which decreases as the square root of the number of galaxies in a pixel. The resulting noise is called shape noise.}, it would be arise from sample variance and be equal to the spectrum squared divided by the number of 
independent modes in the bin. In that case, since the number of modes in a bin grows as $l$ 
for log binning, the $(S/N)^2$ would grow as $\lmax^2$. 

Fig.~\rf{sn} shows the $(S/N)^2$ as a function of $l_{\rm max}$. 
 The linear $\kappa$ field is shown by the dotted gray line. The information obtained from the
nonlinear $\kappa$ field falls well below this ideal limit, as seen in
the figure. This  arises because the nonlinearities
significantly affect the covariance
matrix. Non-zero off-diagonal elements in the covariance matrix mean that many of the modes carry redundant
information, so the total gain is significantly below the $\lmax^2$
Gaussian limit. The log transform undoes a large portion of this damage. 
The left panel of 
Fig.~\rf{sn} shows that the information in $\knew$ is well above that in $\kappa$ and  close 
to the Gaussian case. 
In other words, we measure 
the amplitude of 
the power spectrum with higher precision if we use the log-transformed field. We find a factor of $\sim 1.3$ improvement in $(S/N)^2$ at $\lmax \sim  250$, a factor of $\sim 2.6$ at $\lmax \sim  1000$, 
a factor of 4 at $\lmax \sim  2000$,
and a factor of 8 at $\lmax \sim 5000$.

\begin{figure}[thbp]
\sfig{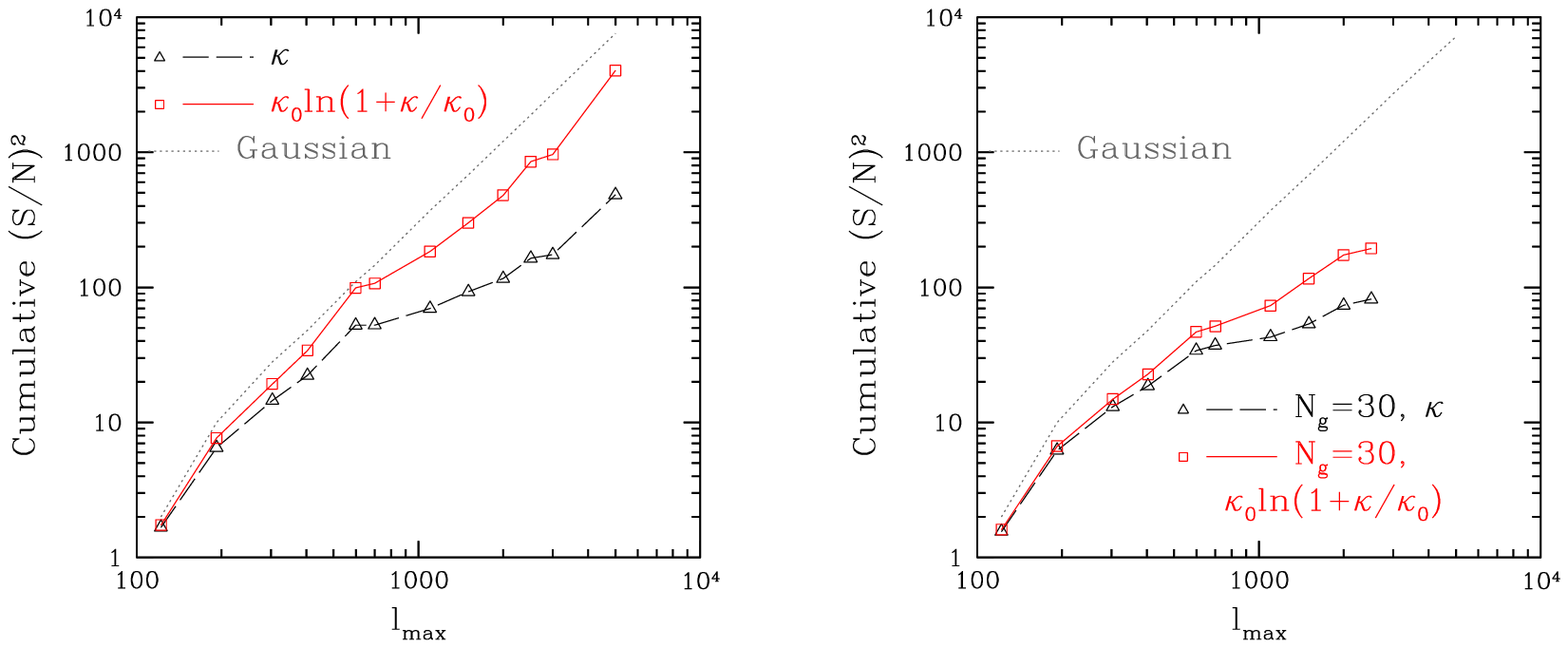}{0.95\columnwidth}
\caption{
Left: the information, represented by $(S/N)^2$, contained in 
the two fields $\kappa$ and $\knew$ in comparison with the Gaussianized field. The
information in the 
 Gaussianized $\kappa$ field (dotted curve) 
increases as $\lmax^2$ as smaller scales are included. The
actual nonlinear convergence $\kappa$ (dashed black line/triangles) loses much of
the $(S/N)^2$ at large $l$, while the log transform (solid red
line/squares) recovers it. 
Right: the
effect of the log transform in the presence of shape noise: we assume a
galaxy number density 
$N_g=30/{\rm arcmin}^2$ at $z_s=1$ and 
increase the pixel size to 
 $2.4 ~ {\rm arcmin}$ 
(accordingly we use $\kappa_0=0.112$). 
We find an improvement of 1.7 (2.4)
in the information content for $\lmax \sim 1000$ (2000)
even in the presence of
 shape noise.  }
\label{fig:sn}
\end{figure}

The restored information in the $\knew$ field can be understood by examining  the covariance
matrix of the power spectra. Fig.~\rf{off} 
shows two rows of the covariance matrix for the fields, with one of the wavenumbers fixed 
at $l'=253$ and $l'=1049$ in the two cases (upper and lower panel). The $\kappa$ covariance matrix has large off-diagonal elements in adjacent bins -- these carry redundant information and therefore do not add much to the $S/N$. The transformed $\knew$, on the other hand, is much more nearly diagonal.
A nearly diagonal covariance matrix implies another important advantage
of the log transform: the approximation of a Gaussian covariance matrix
for cosmological parameter estimation is more accurate 
for $\knew$.
\Sfig{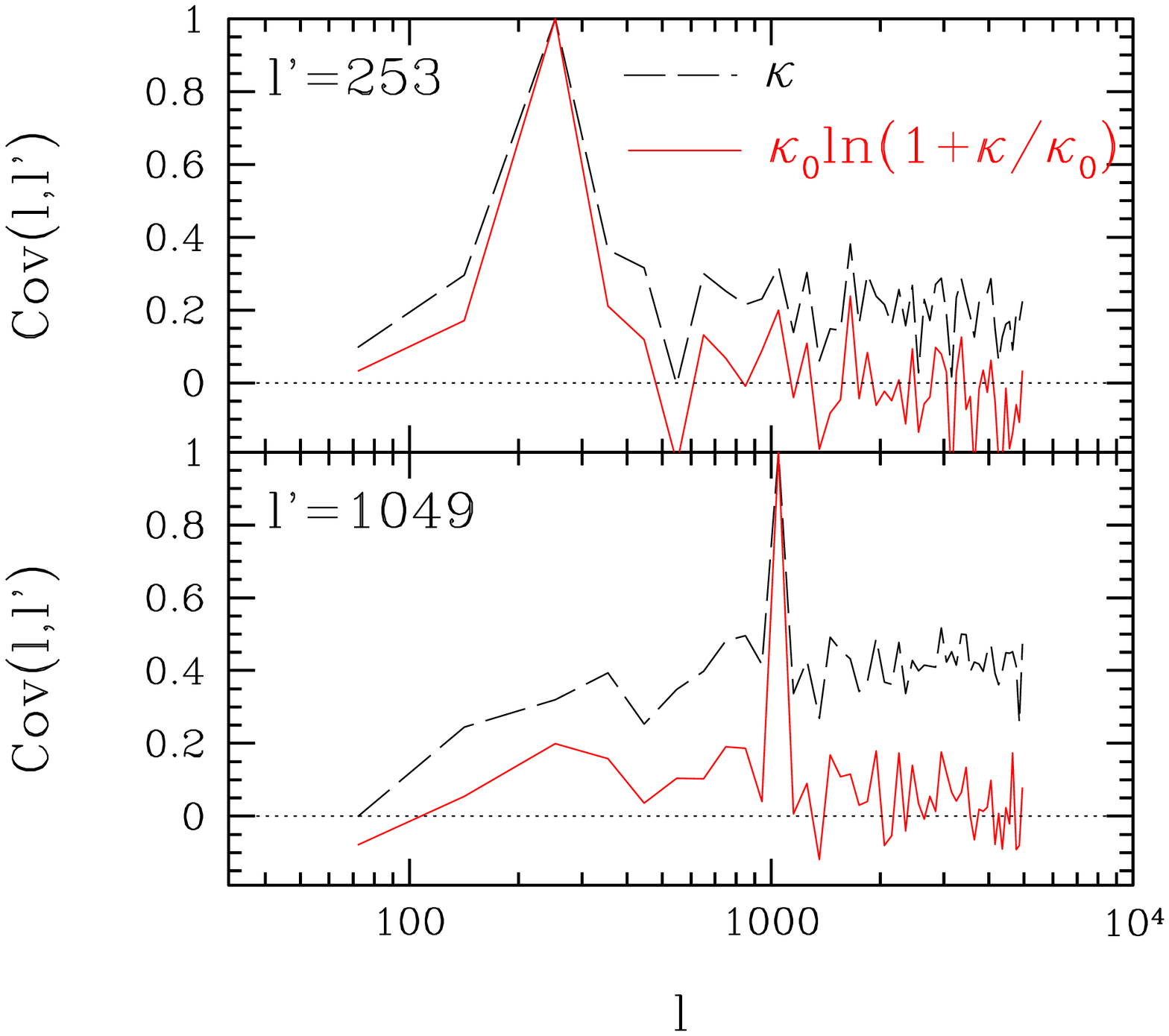}{Slices of the covariance matrix. 
The off-diagonal elements are normalized relative to the diagonal elements, i.e., $\cov(l,l')/(\cov(l,l)\cov(l',l'))^{1/2}$ are shown as a function of $l$ for two choices of $l'$. The off-diagonal covariances between different scales have been substantially decreased by the log transform. }

Another way of understanding the gain in information in the log field is
to consider the Taylor expansion of 
the log transform $\knew$. 
For $-1< \kappa/\kappa_0 \le 1 $,
 one sees that $\knew$ contains the
standard convergence field, but also a piece that scales as
$\kappa^2$ (and higher orders). Considered perturbatively, then, the spectrum of $\knew$
will depend not only on the 2-point function of $\kappa$, $C_l$, but
also on the 3-point function, the bispectrum, as well as higher-point
functions. Effectively, then this rather simple transform captures
information in the power spectrum, bi-spectrum, tri-spectrum,
etc. in a compact way. 
 Of course, it does not contain all the
information in these higher point functions, but the improvement seen in
Fig.~\rf{sn} suggests that using $\knew$ as 
a transform in future
surveys may be a simple, powerful way to bundle much of this information
into one simple spectrum. 
We have tested this by measuring the information contained in $\knew'\equiv \kappa -\kappa^2/(2\kappa_0)$ and found that, once we apply an appropriate cutoff\footnote{We remove the high $\kappa$ values by replacing $\kappa$ larger than 0.1 with 0.1.} on high $\kappa$ values to make the polynomial expansion more sensible,
the single extra term replicates most of the improvement observed in the log transform.
Meanwhile the cross-correlation of the $\kappa$ and
$\kappa-\kappa^2/\kappa_0$ which involves only up
to bispectrum, with an appropriate cutoff, replicates most of the
improvement 
up to $l\sim 1000$. This implies that the bispectrum is the dominant contributor to this improvement up to the scales.

There are several caveats to this analysis. So far, we have neglected noise, in particular shape noise due to the
random orientations of galaxies on the sky. We have studied this issue
for several survey parameters. Surveys with higher number density have lower shape noise and
therefore the advantages of $\knew$ approach those depicted in 
the left panel of Fig.~\rf{sn}.
For a galaxy number density of 30 per square arcminutes at $z_s=1$, 
as expected for the
planned Subaru Hyper SuprimeCam survey~\cite{2006SPIE.6269E...9M},
we find an improvement of 1.7 (2.4) 
in the information content for $\lmax \sim 1000$ (2000) (right panel in
Fig.~\rf{sn}). The gain is larger for more ambitious surveys like LSST or DUNE~\cite{Tyson:2006hs,Refregier:2008js}
and smaller for  shallower surveys like the Dark Energy Survey~\cite{Abbott:2005bi}. 

Second, although $\knew$ has some of the advantages of the linear $\kappa$ field, it does not actually recover the initial field phase by phase
since the cross-correlation between the initial and final fields,
when tested for the density fields,
 does not improve by this transformation. 
Third, our analysis (and our definition of {\it information}) revolved around only one parameter, the amplitude of the 
power spectrum. Its shape and evolution certainly contain 
additional cosmological information, as discussed by \cite{TakadaJain:09}. 
Finally, we
have assumed that the convergence field, reconstructed from the shear,
will be available over the entire survey area -- in practice such a
reconstruction adds additional noise. 
We are in the process of studying these issues, but they are not
expected to affects our main point:
that the log transform $\knew$ recovers cosmological information.

\section{Conclusion}
We have found that the log transform of Eq. (1) alters the nonlinear
convergence field to one that mimics the properties of a Gaussian
field. It returns a PDF that is close to a Gaussian -- analogous to the
findings of \cite{Neyrinck:2009fs} for 
the 3D density field. The
signal-to-noise (i.e., precision) of the measurement of the amplitude
of the power spectrum is greatly improved over a wide range of angular scales, $200\lsim l\lsim 10^4$.
Even in the presence of shape noise, this improvement holds, to a
greater or lesser extent depending on the galaxy number density. 
The improvement arises from the effect on the covariance matrix: 
the off-diagonal elements of the covariance matrix are 
substantially reduced for the log transform. 
We find that the bispectrum that is embedded in the log 
transform
is the dominant contributor to this improvement.
Therefore the log transform appears to bundle much of the information
from higher order statistics into the power spectrum.

Upcoming imaging surveys will collect data on the shapes of galaxies at
an unprecedented rate,  with an eye towards understanding  
the physics which drives the acceleration of the Universe. 
It is imperative that we use algorithms to analyze this data which
extract as much of the cosmological information as possible: the log
transform $\knew$ is a step in this direction.

\bigskip
\noindent{\bf Acknowledgement}: 
We would like to thank David Johnston, Nick Gnedin,  Mark Neyrinck 
and Alex Szalay for useful comments. 

This work is supported in part by World Premier
International Research Center Initiative (WPI Initiative), by
Grant-in-Aid for Scientific Research on Priority Areas No. 467 ``Probing
the Dark Energy through an Extremely Wide and Deep Survey with Subaru
Telescope'', MEXT, Japan, by the JSPS Research Fellows, by the DOE at Fermilab through grant DE-FG02-95ER40896, and by NSF
Grants AST-0908072 and AST-0607667. 

\bibliography{lensing}
%%%%%%%%%%%%%%%%%%%%%%%%%%%%%%%%%%%%%%%%
%%%%%%%%%%%%%%%%%%%%%%%%%%%%%%%%%%%%%%%%
\end{document}